DPHEP-2009-001
November 30, 2009# Data Preservation in High-Energy Physics

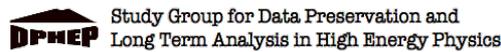

Study Group for Data Preservation and
Long Term Analysis in High Energy Physics

http://dphep.org

**Abstract**

Data from high-energy physics (HEP) experiments are collected with significant financial and human effort and are mostly unique. At the same time, HEP has no coherent strategy for data preservation and re-use. An inter-experimental Study Group on HEP data preservation and long-term analysis was convened at the end of 2008 and held two workshops, at DESY (January 2009) and SLAC (May 2009). This document is an intermediate report to the International Committee for Future Accelerators (ICFA) of the reflections of this Study Group.1

# Executive Summary

Large data sets accumulated during many years of detector operation at particle accelerators are the heritage of experimental high-energy physics (HEP). These data sets offer unique opportunities for future scientific studies, sometimes long after the shut-down of the actual experiments: new theoretical input; new experimental results and analysis techniques; the quest for high-sensitivity combined analyses; the necessity of cross checks. In many cases, HEP data sets are unique; they cannot and most likely will not be superseded by data from newer generations of experiments. Once lost, or in an unusable state, HEP data samples cannot be reasonably recovered. The cost of conserving this heritage through a collaborative, target-oriented long-term data preservation program would be small, compared to the costs of past experimental projects or to the efforts to re-do experiments. However, this cost is not negligible, especially for collaborations close or past their end-date.

The preservation of HEP data would provide today's collaborations with a secure way to complete their data analysis and enable them to seize new scientific opportunities in the coming years. The HEP community will benefit from preserved data samples through re-analysis, combination, education and outreach. Funding agencies would receive more scientific return, and a positive image, from their initial investment leading to the production and the first analysis of preserved data.

An international Study Group reviewed the current state of data preservation in HEP: summarised physics arguments for preservation, outlined possible data preservation models, described technologies and facilities and examined the aspects of supervising and governing data preservation. The preliminary conclusions of the Study Group are:
1. Data preservation beyond the end-date of experiments opens up future scientific opportunities. Given the present status of experimental programs at most facilities, an urgent and vigorous action is needed to ensure data preservation in HEP.
2. Different levels of data preservation and usability are possible. The preservation of the full analysis capability of experiments is recommended, including the preservation of reconstruction and simulation software. A dedicated project in each experiment is needed to assess the corresponding technological requirements.
3. The technological aspects of data preservation are well within the reach of large computing centres in HEP. Nevertheless, an interface to the experiment know-how should be introduced. The most efficient solution would be the creation of a data archivist position, in charge with the preservation of the data analysis capabilities.
4. The preservation of HEP data requires a synergic action of all stakeholders: experimental collaborations, laboratories and funding agencies. A clear and internationally coherent policy should be defined and implemented.
5. An International Data Preservation Forum is proposed as a reference organisation, with the mandate to organise and overview HEP data preservation initiatives; to discuss and propose solutions to technological or policy issues; to evolve into a "clearing house" for policies for access and re-use of preserved data. The Forum should represent experimental collaborations, laboratories and computing centres.

# Introduction

High-energy physics (HEP) experimental collaborations summarise their scientific results through publications in peer-reviewed journals or conference proceedings. Long-term preservation and re-use of the primary data beyond the published analyses is generally not pursued in HEP. Experimental collaborations maintain data-analysis capability for some time after the end of the data taking, on average five to ten years. In most cases, with at least one notable exception to date, the basic (raw) data disappear after that period. The main reasons are rapid changes in storage technologies and, especially, computing and software systems, which are not matched by an effort to migrate the data-analysis infrastructure and to maintain the expertise level needed for new analyses. Reports about data from the LEP experiments being increasingly difficulty to analyse or even becoming inaccessible are starting to surface.

The scientific value of long-term analysis of HEP data is difficult to underestimate and well understood by the HEP community: a recent survey by the PARSE-Insight project[1] found that out of over a thousand HEP physicists, around 70% see data preservation as "very important" or even "crucial". Given that experiments such as those at HERA or B-factories are entering their final analysis period, immediate action should be taken.

In order to investigate the opportunities and requirements of HEP data preservation a Study Group on HEP data preservation and long-term analysis was formed[2] at the end of 2008 and held two workshops, in DESY in January 2009 and in SLAC in May 2009. This Study Group is organised in four working groups, dedicated to various aspects of HEP data preservation: the physics case; preservation models; preservation technology; governance. The following chapters reflect the preliminary outcome of these four working groups.

# The Physics Case for Data Preservation in HEP

Long term preservation of HEP data is crucial to preserve the ability of addressing a wide range of scientific challenges and questions at times long after the completion of experiments that collected the data. In many cases, these data are and will continue to be unique in their energy range, process dynamics and experimental techniques. New, improved and refined scientific questions may require a re-analysis of such data sets. Some scientific opportunities for data preservation are summarised in the following points:

**Long-term completion and extension of scientific programs**
This entails the natural continuation of the physics program of the individual experiments, although at a slower pace, to ensure a full exploitation of the physics potential of the data, at a time when the strength of the collaboration (analyst person-power as well as internal organisation) has diminished. It is estimated that the scientific output gained by the possibility to maintain long-term analysis capabilities represents roughly 5 to 10% of the total scientific

---

[1] Andre Holzner, Peter Igo-Kemenes, Salvatore Mele . "First results from the PARSE.Insight project: HEP survey on data preservation, re-use and (open) access", CERN-OPEN-2009-006, Jun 2009, e-Print: arXiv:0906.0485 [cs.DL]
[2] The composition of the Study Group is detailed in Appendix A

production of the collaborations. More important than the sheer number of publications is the nature of these additional analyses. Typically, these analyses are the most sophisticated and benefit from the entire statistical power of the data as well as the most precise data reprocessing and control of systematic effects.

**Cross-collaboration analyses**
The comprehensive analysis of data from several experiments at once opens appealing scientific opportunities to either reduce statistical and/or systematic uncertainties of single experiments, or to permit entirely new analyses which would be otherwise impossible. Indeed, ground-breaking combinations of experimental results have been performed at LEP, HERA and the TeVatron, during the collaborations' lifetime, providing new insight in precision measurements of fundamental quantities, and extending the ranges for search of new physics. Preserved data sets may further enhance the physics potential of experimental programs, by offering the possibility of combinations which would not be otherwise possible. Data from facilities where no active collaboration is operating would be available for combination with new data. At the same time, well-documented preserved data would also enhance opportunities for combinations among current experiments, which may be otherwise prevented by the lack of standards leading to insurmountable technical or scientific problems. The HEP community comprises sub-communities of experts in various fields such as flavour physics, neutrino physics, and so on. These expert communities would greatly benefit from having simultaneous access to data sets from relevant experiments. For example, B-physics experts could devise analyses simultaneously using data from BaBar, Belle, Cleo-C. Such an effort to combine analyses is already ongoing, for example between the H1 and ZEUS collaborations, and an evaluation of such an approach is underway between the Belle and BaBar collaborations. An effort in standardising and/or documenting data sets for long-term preservation would have an immediate return in facilitating these combinations.

**Data re-use**
Several scientific opportunities could be seized by re-using data from past experiments. For instance, new theoretical developments could allow new analyses leading to a significant increase in precision for the determination of physical observables. Theoretical progress can also lead to new predictions (e.g. of new physics effects) that were not probed when an experiment was running and is not accessible at present-day facilities. Similarly, new experimental insights (e.g. breakthroughs in Monte Carlo simulation of detector response) or new analysis techniques (e.g. multivariate analysis tools, greater computing capabilities) could allow improved analyses of preserved data, with a potential well beyond the one of the published analyses. Results at future experimental facilities may require a re-analysis of preserved data (e.g. because of inconsistent determinations of physical observables, or observation of new phenomena which may/should have been observed before). For example results from the LHC experiments may very well induce re-analysis of LEP, Tevatron or HERA data.

**Education, training and outreach**
Preserving data opens new opportunities in training, education, and outreach. It permits data analysis by undergraduate or graduate students, without restriction to institutes that collaborated to the experiments, opening new opportunities for institutes in developing countries to initiate and develop HEP research. The benefit to the field is the ability to attract and train the best inquisitive minds. It also gives unprecedented opportunities to teach hands-on classes in particle physics, experimental techniques, statistics, and to explore physics topics that would not have been otherwise covered. High schools students could be exposed to simplified and highly visual analyses (similar to the successful EPPOG[3] master classes using which use special sub-sets of the DELPHI and OPAL data), in order to re-ignite the general public interest in the field and to attract new students to physics.

**An example of successful reanalysis of HEP data**
The possibility offered by long term data preservation is best illustrated by the recent resurrection and re-analysis of data from JADE, an experiment that operated at the PETRA $e^+e^-$ collider between 1979 and 1986. Applying new theoretical input and new experimental insights and methods, the old data provided new results on the QCD coupling $\alpha_s$ and its energy dependence, in an energy range which today is not otherwise accessible, and also allowed combined analyses with data from OPAL at LEP[4]. This re-analysis of data that is more than twenty years old was made possible by the commitment of a few individuals. It has been a tour de force and far from a standard enterprise in HEP. The analysis of this example shows that the preservation of HEP data at the highest level can be successful in the presence of proper means.

# Models for HEP Data Preservation

Different preservation models can be organised in levels of increased complexity. Each level is associated with one or more use cases. The preservation model of an experiment should reflect the level of the use cases to be enabled in the future, and the whole aim of the preservation exercise. A survey of a few computing models revealed that the amount of data (including simulated data) of current experiments is between 0.5 PB and 10 PB, which is a significant but manageable size[5]. The costs related to maintaining and migrating the software and the data analysis infrastructures, to effectively preserve them, are model dependent and are difficult to estimate. Nevertheless, the cost of various preservation models is expected to be primarily driven by person power requirements rather than the cost of data storage.

Different preservation models are summarised in Table 1 and presented in the following, with remarks on the associated cost estimates and benefits. The implementation of these models at the beginning of the lifetime of an experiment will greatly increase the likelihood of success, minimise the effort and ease the use of the data in the final years of the collaborations.

---

[3] The European Particle Physics Outreach Group, http://eppog.web.cern.ch/eppog.
[4] See for example: JADE Collaboration (S. Bethke et al.). MPP-2008-131, Oct 2008. 9pp. Submitted to Eur.Phys.J.C. e-Print: arXiv:0810.1389 [hep-ex]
[5] 1 PetaByte (PB) = $10^{15}$ bytes; for comparison, the four LHC experiments will produce about 15 PB of data per year and Google computing centres process 20 PB per day.

| **Preservation Model** | **Use case** |
| --- | --- |
| 1. Provide additional documentation | Publication-related information search |
| 2. Preserve the data in a simplified format | Outreach, simple training analyses |
| 3. Preserve the analysis level software and data format | Full scientific analysis based on existing reconstruction |
| 4. Preserve the reconstruction and simulation software and basic level data | Full potential of the experimental data |

Table 1: Various preservation models, listed in order of increasing complexity. Subsequent models are inclusive. For example, preservation model 4 also includes steps and use cases described in models 1,2 and 3.

**Level 1: Provide additional documentation**
A model of preservation, without actually preserving the data, would be to provide additional documentation. Such a practice is also a recommendation to any preservation effort, and as such the guidelines in this section apply to all models. Additional documentation may include: more information associated with, or embedded in, publications (extra data tables, high-level analysis code, etc.); internal collaboration notes; meta-data related to the running conditions; technical drawings; general experimental studies (for example on systematic correlations); an expert information database (for instance minutes, slides, news); documents available on paper only that could be digitised and stored in electronic format. Care should be taken to remove or tag the redundant or noisy information which often appears during the analysis (for instance intermediate, non-validated hypothesis or non-pursued technical solutions may appear in the daily exchanges but be irrelevant for the final analysis configuration).

In the process of the documentation preparation, global information infrastructures in the community as well as those within experimental collaborations may be beneficial for a robust preservation project. An organised internal documentation migration to a HEP community information system like INSPIRE[6] would be one way to achieve this goal. Auto-documentation tools like those included in ROOT[7] should be used to their maximum ability. Day-to-day documentation within the collaboration may be stored in a wiki which also has the advantage of simple (text-like) preservation option. A common format for popular tools (e.g. electronic log books) would also be useful, enabling such metadata to be preserved in a similar way. It would be beneficial to experiments to consult with a professional archivist who is aware of the standards within the HEP community and elsewhere. In particular, the proper storage and possible digitisation of paper documents should be pursued in collaboration with libraries.

---

[6] INSPIRE is the new information platform for HEP, realised by CERN, DESY, FERMILAB and SLAC, which will replace and enhance the popular SPIRES system. http://www.projecthepinspire.net
[7] ROOT is the analysis software framework based on the C++ programming language (http://root.cern.ch), widely used in high physics analyses, in particular at the LHC experiments.

For a new experiment, the costs of a more coherent, centralised and preservation-oriented documentation strategy would be minimal, if applied from the beginning, whereas the benefits for future use are clearly significant.

To preserve nothing beyond the publications and the associated, improved documentation may be an option only if the belief is that the data are no longer of any scientific use, such that they have been superseded by a new experiment or the full potential of the physics program has been extracted. Past experience demonstrates that this is rarely the case, and concrete examples of scientific benefits of data preservation are given in the previous section. An effort to preserve lower level information needed for a full analysis certainly provides an added value to the scientific reach of an experiment.

**Level 2: Preserve the data in a simplified format**
An economic means of preserving the real and simulated data without the need for any experiment-specific software would be to just preserve the basic, event-level, four-vectors describing the detected particles. This should be done in as simple a structure as possible in order to facilitate future interpretation and re-use of this data. A simple four-vector format can be very useful in terms of a model for outreach and education purposes. However, this format will in general not be sufficient to perform a full physics analysis, except for a few particular cases. It is likely that a dedicated effort would be needed in each collaboration to decide on the physics content of the data format. Further options, like preserving the capability to perform a simplified error propagation, may also be implemented. In terms of the required person-power, this option would require a dedicated effort of the experts to define the relevant information, with a relatively simple technical implementation and modest requirements for long-term maintenance.

**Level 3: Preserve the analysis level software and data format**
This option includes the preservation of analysis level software including non-experiment specific software, such as ROOT. With respect to level 2 this introduces a supplementary dependence on the longevity of the required experiment-specific software and may require a thorough study of the computing environement. An example of problematic dependence is the commercial software (the usual fragility is associated with databases) or domain specific software with limited lifetime (for instance the scientific software library CERNLIB is no longer maintained since more than ten years but is still widely used). More effort than level 2 would most likely be required for the preparation and maintenance of this dataset, especially if backwards compatibility issues arise. However, the benefits of this level of preservation are the ease of analysis and access to extra features and improvements from the software. This option may be sufficient to perform complete analyses when the existing detector and simulated data sets are sufficient for the pursued goal.

**Level 4: Preserve the reconstruction and simulation software and basic level data**
Certain analyses may require the production of new simulated signals or even require a re-reconstruction of the real and/or simulated data. For these, the full reconstruction and simulation software would need to be preserved. This may or may not require basic (raw) level data, depending on what is stored at the more abstract level (usually called DST), which is experiment specific. Generally for greater flexibility all data should be preserved. By preserving the full analysis chain, one retains the ability to derive associated corrections, studies of efficiencies and acceptances, and to perform a full systematic error analysis. Special care should be given to the protection of the sensitive components (official calibrations, simulation tuning etc.) that should not be redone unless high level experts are involved. At this level of preservation the aim is not for a common format but rather a common standard. Significant resources will be needed for this preservation model during the preparation (R&D) and maintenance (archived data) phases. However the clear benefit of such a model is that the full physics analysis chain is available and full flexibility is retained for future use.

**Preparing a data preservation project**
The preservation model should be taken into account as early as possible in the computing strategy, such that the transition to the archival phase is done with a reduced effort. To maximise the efficiency of the preservation project, a collaboration should employ as much centralised software as possible. This also benefits the collaboration by fostering the adoption of common code and results in a more efficient use of person-power. It is likely that the experiments close to the end of the data analysis need a dedicated effort to achieve a reliable implementation of their chosen preservation model. A necessary component of any preservation project is the implementation of robust validation procedures, which should be able to indicate the status of the data analysis capabilities without physics expert intervention. The validation software should be seen as an essential component for the preparation of the technological steps like storage upgrades or operating system migrations, which are the critical moments of a data preservation project.

The R&D effort of the data preservation project should have significant overlap with the collaboration lifetime and should benefit, in addition to the dedicated human resources, from the general expertise in data analysis within the collaboration. In the longer term, the preservation project should be taken as a permanent activity, implemented in the associated host laboratories or computing centres, aimed at maintaining and optimising the exploitation of the preserved data analysis facility. A schematic view of the transition from a full analysis environment to an analysis based on the preservation model (archival phase) is presented in figure 1. The aspects related to the data supervision (including issues of Open Access) are discussed in the last chapter.

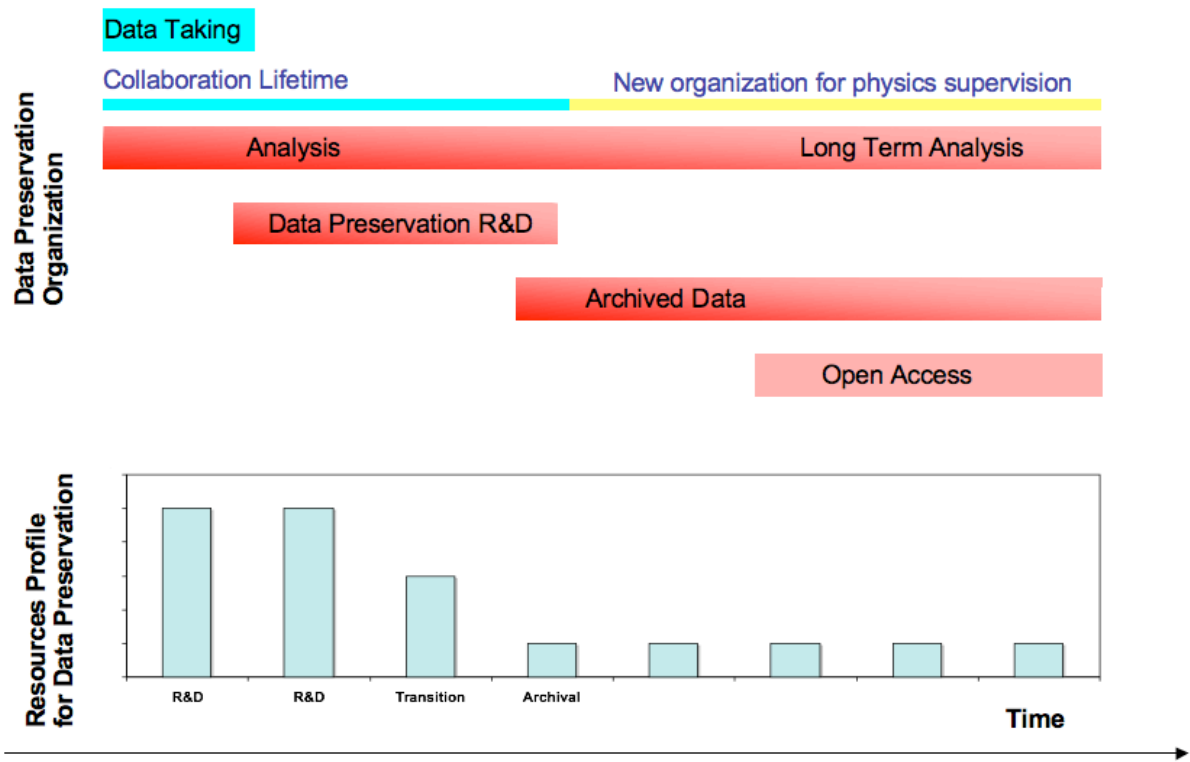

Figure 1: A possible model for data preservation organisation and resources presented as the milestones of the organisation and the resources evolution as a function of time.

## Technologies and Facilities

Support for data storage, data access, analysis computing, networking, code management, and related facilities are required during the various phases of an experiment during and after the end of data taking. There are efforts and costs associated with these that must be identified and funded.

**Effort and costs to facilities**
Some specific costs that can be identified and accounted for include the following. There are costs to transition from the *analysis* to the *archival* phase. This includes database reconfiguration, documentation, data copying, operating system upgrades and so on. Beyond the archival phase there are costs associated with the long-term support of preservation services, including access and regular data validation. Data validation is an important function that can help ensure that the use of the preserved data and associated systems and software are able to provide reliable and reproducible results as the facilities are upgraded and modified over time.

**Funding and agreements**
Funding arrangements for the facilities should be based on a well-defined model of data preservation and analysis (analysis phase, transition phase, archival phase, Open Access phase, and so on). The size of the resources required for each phase should be estimated and agreed to by the experiment, the facility or facilities involved, and the funding bodies.

Minimisation of the data volume and complexity is useful in minimising costs during the archival phase. There is a trade-off between the effort required to minimise the data volume or other simplifications and the benefits that come from doing so, and these should be recognised.

Consistent agreements for support are recommended and should be developed among the experiment, the facility or facilities involved and the funding agencies. These agreements should cover all phases and types of access.

Possible long-term Open Access to data, potentially including education and outreach, should be recognised as a key part of the process, which is likely to require additional or different support and infrastructure from that provided in the archive phase alone.

**Archival expertise**
We recommend that expert assistance should be funded as part of long-term data preservation and analysis. This could be an archivist[8] position filled by a physicist: a new type of position created specifically to address issues related to long-term data analysis and preservation. This post is likely to be similar to those found in existing data centres for organisations such as the British Atmospheric Data Centre (funded by the National Environmental Research Council in the UK). Experts in their discipline, who understand the science and its current issues, are responsible for validating and maintaining the data, and ensuring its integrity as it is migrated to new technologies during the archive phase. In addition, they liaise with the users, advising them about data availability, data quality, and appropriateness for the user's needs. The amount of effort required for this work will vary but has a component directly proportional to the number and type of end users accessing the data. The archivists are themselves often also leading on-going research in one or more related areas of the discipline.

**Technologies**
Technology investigations should be part of this process. The focus should be on minimising the complexity of the system and long-term support needs. However, it is acknowledged that technological change will occur, and thus generate the need for modifications and upgrades in the future. Virtualisation is a promising technology for encapsulating a stable system for use over long periods of time. This should be investigated and tested as part of the work required to implement the proposed systems.

# Governance and Supervision of Data Preservation

The data preservation process should follow well-defined policies, defined as soon as possible during the lifetime of the collaborations, and possibly embedded in a global HEP data-preservation initiative. Preservation policies should address the aspects discussed in this document: the physics case, the preservation model and the technological aspects. In addition, it should address the following items:

---

[8] The Oxford English Dictionary, fittingly, defines an "archivist" as "the person who maintains and is in charge of archives", in turn defined as "a collection of historical documents and records [… and] the place where such documents or records are kept."

**Supervision of the data preservation process**

Data preservation is likely to include complex technical aspects and can be affected by their time profile: intense activity corresponding to major technological operation (e.g. changes in storage media or operating systems), or periods with reduced activity, with a risk of gradual dispersion of the know-how. The data archivists are likely to constitute a minority in the participating computing centres and in some cases they may be partially allocated to other tasks. The organisation of the preservation process should therefore include supervision mechanisms aimed at enforcing the contracts between the collaborations and the computing centres (facilities) and to make sure that the necessary level of expertise is maintained. The technical actions defined in the preservation model should be constantly reviewed.

**Access to preserved data**

The key motivation for the long-term preservation of HEP data is the unique scientific opportunities opened up by their re-use. In some scientific cases this re-use is by members of the collaborations who originally took the data. In other scientific cases, new opportunities are generated by the re-use of data by scientists not originally involved in the collaboration. Eventually, Open Access could generate further opportunities to use preserved HEP data. The PARSE.Insight study of over a thousand HEP scientists found that 54% of the theorists and 44% of experimentalists think that access to data from past experiments could have improved their scientific results. While Open Access to preserved HEP data, immediate or at a later stage, generates new opportunities, it also raises new issues in our community, such as the scientific responsibility for results obtained from preserved data sets. The survey also found that 45% of the respondents are "very concerned" or "gravely concerned" that re-use of data may in general lead to an inflation of incorrect results. At the same time, as many as 53% of the respondents are concerned about incorrect results due to a misinterpretation of the preserved data.

Open Access to data, albeit of often vastly larger simplicity, is sometimes the norm in other disciplines. HEP colleagues who started programs in ground or satellite-based astro-particle physics have met and adapted to these different realities. The opportunities held by Open Access to preserved data have to be evaluated against the concerns that are raised, so that informed policy decisions can be made at the highest level. These considerations should be decoupled from pursuing the unavoidable and necessary steps in data preservation, and addressed through a parallel, wider debate.

**Physics supervision**

The publication of physics results during the lifetime of a collaboration follows rigorous procedures, exercised over many years. In order to ensure a proper usage of the preserved data, a scrutiny process of the physics output obtained from this data should be defined. Certification mechanisms ensuring the correctness of the produced results should be therefore implemented, reflecting the quality requirements specific to the level of detail used in the analysis.

**Authorship**

The author lists of the HEP publications are defined according to internal mechanisms and include usually all members of the collaboration. Beyond the lifetime of a collaboration, the authorship rules for the scientific papers obtained using the preserved data sets should be clearly defined such that data analysis is encouraged and the proper credits are allocated to the collaboration that collected the data. The authorship rules should be linked to the physics supervision process.

**Outreach and education**

The preserved data contains in many cases examples of basic subjects in the high energy physics history and research. These textbook examples can be used with high efficiency for educational activities and also to communicate the basics of the discipline to a large audience.

**Definition and endorsement**

The preservation of HEP data requires a synergic action of all stakeholders: experimental collaborations, laboratories and funding agencies. A clear and internationally coherent policy should be defined, endorsed and implemented at the following levels:
- The running experiments should formulate their data preservation strategy
- The HEP laboratories involved as host laboratories and data centres should state their data preservation strategy
- The funding agencies should endorse a clear and coherent policy of data preservation for HEP experiments

**Global solutions**

Due to the international nature of the HEP research, data preservation is a global issue and should be a treated in a global way. An International Data Preservation Forum is proposed as a reference organisation, with the mandate to organise and overview HEP data preservation initiatives; to indicate solutions to technological or policy issues; to evolve into a "clearing house" for policies for access and re-use of preserved data. The Forum should represent experimental collaborations, laboratories and computing centres. It can be structured around a lightweight organisation composed of a Study Group chaired by a coordinator, supported by an assistant. The experiments and computing centres are represented in the Steering Committee. The Advisory Committee is composed of independent personalities. The central management reports to the funding agencies and to international bodies such as ICFA. A possible organisational scheme is presented in figure 2. The role of the forum can extend from data preservation policies and archival systems overview to scientific communication and outreach activities. It can give the opportunity for coordinated international projects in scientific and technological developments related to the archived data.

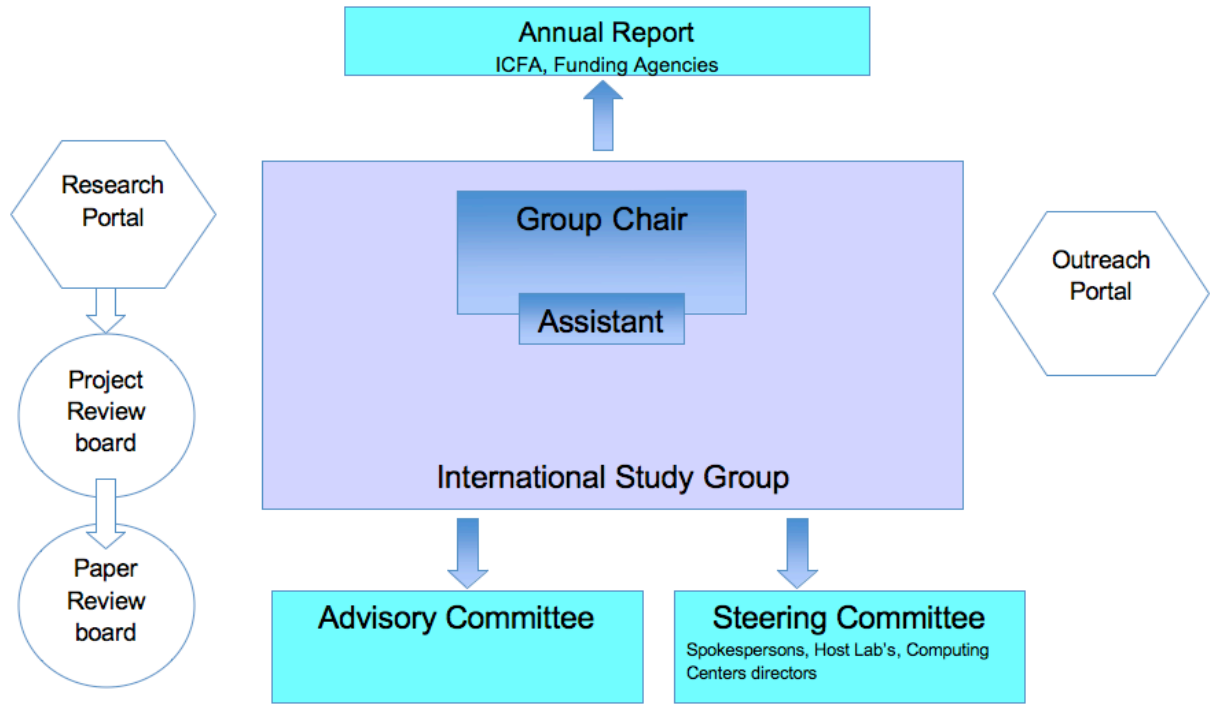

Figure 2: Organisation of an international forum for data preservation in high-energy physics. The group is structured around a chair and his assistant, in charge with the communication with the steering and advisory committees and with ICFA. This lightweight structure is already informally in place. The research and the outreach portals are possible longer term extensions of the present activities.

## Conclusions

Large data sets accumulated through many years of detector operation at particle accelerators are the heritage of experimental high-energy physics (HEP), but no coherent efforts are taken to preserve this data. The cost of a target-oriented long-term data preservation program would be small when compared to the initial costs of past experimental projects or to the efforts to re-do experiments. A Study Group on HEP data preservation and long-term analysis reviewed the current status of data preservation in HEP; summarised physics arguments for preservation; outlined possible data preservation models; described technologies and facilities; examined a structure for governance of data preservation. The preliminary conclusions are:
1. Data preservation beyond the end-date of experiments is vital to seize future scientific opportunities and an urgent and vigorous action is needed.
2. Different levels of data preservation and usability are possible, from enhanced documentation full analysis capability. The latter is recommended to fully address future scientific challenges. A dedicated R&D project in each experiment is needed to assess the corresponding technological requirements.

3. The technological aspects of data preservation are well within the reach of large computing centres of the field, which need to be complemented by the appropriate know-how in the newly-created position of "data archivist".
4. The preservation of HEP data requires a synergic action of all stakeholders: experimental collaborations, laboratories and funding agencies. A clear and internationally coherent policy should be defined and implemented.
5. An International Data Preservation Forum is proposed as a reference organisation, to organise and overview HEP data preservation.

# Outlook

The Study Group has already focussed on a number of open points and plans to continue their evaluation in the next step. The reflections contained in the main body of this document were submitted to the Advisory Committee and to ICFA during the second half of 2009. The initiative was positively received and the Study Group, endorsed by ICFA since August 2009, was encouraged to continue the common work, towards the definition of an International Organisation as a means to obtain a coherent approach of data preservation in HEP. This section summarises the main ideas for further working directions, the majority of which were inspired by these discussions.

More examples of past experiences should be documented, for example the recent publications with LEP data ($\alpha_s$ at NNNLO) or similar examples from DIS experiments. Not only positive but also negative cases may have an instructive value. Some open cases involving inconsistencies between the old data and recent measurements may have been solved with adequate access to previous experimental data. Physics cases can certainly already be identified through the present publication and should be evaluated in more detail. In particular, "simulations" of data re-analysis should be attempted such that the necessary ingredients can be identified and the required resources better evaluated. In particular, the timescales for preservation and the targeted precision of the analyses expected to be performed on the preserved data sets should be associated with specific data preservation models.

The differences between preservation levels 1 to 4 should be sharpened, with target-oriented strategies. Concerning the highest abstraction level, a few ideas have been proposed, including a "student's edition" with high quality subsets for typical analyses, featuring simplified structure and access. A more complex level on clean samples, to be used for training of Ph.D students in parallel with their research subject can be investigated as a further step. Retaining the full analysis capability represents a major enterprise and an step-by-step approach could be considered in order to asses the requirements for the full preservation.

Standardisation was recommended as a valuable working direction, concerning both the data formats, the technological transitions and also the software engineering. In this three-way context, the validation process related to the preservation process is a major line of investigation. Standardisation can be used as a working tool towards improving the scientific coverage of more abstract data layers. Virtualisation is a promising approach which should be carefully considered in the context of the recent evolution towards cloud computing.

The group should be more determined in asking for contemporary documentation of experimental results in high energy physics in the public domain: letters accompaned by public notes, supporting documents attached to papers, and the release of all documents even in unedited form. This action, even for running or starting experiments, is also recommended by the IUPAP C11 proposals for using notes as an external documentation and recognition support.

The physics supervision of the preserved data sets was stressed as an important component of the effort. It is related with expertise that should be retained around the preserved data sets, in order to ensure a correct and efficient access to the data. It is recognised already by the Study Group that this crucial issue should be taken into account and concrete algorithms should be investigated towards a functional and reliable model for access to the preserved data. In particular, authorship and review procedures should be investigated and proposed.

In the absence of an already installed practice for long term organisation, data from large experiments are in danger. The rapid change in funding profiles for HEP within labs may also endanger data samples. Data preservation projects should be guaranteed and independent of such decisions. Immediate costing and first step measures should be taken to avoid irreversible actions.

A more involved comparison with other domains, in particular with astrophysics has begun and will be pursued further. Other domains may be relevant: meteorology, climatology, astronomy as well as those in the field of humanities, all rely on access to large quantities of unique data. Such a comparison may lead to new ideas on inter-experiment cooperation, standards and data access, and is likely to be fruitful for shaping the HEP data preservation initiative.

# Appendix A: The ICHFA DPHEP International Study Group
Members of the International Study Group

| Name | Email | Affiliation |
|---|---|---|
| Richard Mount | richard.mount@slac.stanford.edu | SLAC, Computing Centre |
| Travis Brooks | travis@slac.stanford.edu | SLAC / SPIRES |
| Francois Le Diberder | diberder@slac.stanford.edu | BaBar, Spokesperson |
| Gregory Dubois-Felsmann | gpdf@slac.stanford.edu | BaBar /LSST DM Syst.Arch. |
| Homer Neal | homer@slac.stanford.edu | BaBar, Computing Coordinator |
| Matt Bellis | bellis@slac.stanford.edu | Babar |
| Amber Boehnlein | amber.boehnlein@science.doe.gov | Fermilab, Computing Division/ D.o.E |
| Margaret Votava | votava@fnal.gov | Fermilab, Computing Division |
| Vicky White | white@fnal.gov | Fermilab, Computing Division |
| Stephen Wolbers | wolbers@fnal.gov | Fermilab, Computing Division |
| Jacobo Konigsberg | konigsb@fnal.gov | CDF, Spokesperson |
| Robert Roser | roser@fnal.gov | CDF, Spokesperson |
| Rick Snider | rs@fnal.gov | CDF, Offline Co-leader |
| Donatella Lucchesi | donatella.lucchesi@pd.infn.it | CDF, Offline Co-leader |
| Dmitri Denisov | denisovd@fnal.gov | DØ, Spokesperson |
| Stefan Soldner-Rembold | soldner@fnal.gov | DØ, Spokesperson |
| Qizhong Li | qzli@fnal.gov | DØ, Co-leader of Computing |
| Erich Varnes | varnes@fnal.gov | DØ, Co-leader of Computing |
| Alan Jonckheere | jonckheere@fnal.gov | DØ |
| Martin Gasthuber | martin.gasthuber@desy.de | DESY, IT Group |
| Volker Gülzow | guelzow@mail.desy.de | DESY, IT Group |
| Yves Kemp | yves.kemp@desy.de | DESY, IT Group |
| Dmitri Ozerov | ozerov@mail.desy.de | DESY, IT Group |
| Cristinel Diaconu | diaconu@mail.desy.de | H1, Spokesperson |
| David South | david.south@desy.de | H1, Software Coordinator |
| Bogdan Lobodzinski | bogdan@mail.desy.de | H1, MC Coordinator |
| Jan Olsson | jan.olsson@desy.de | H1/JADE |
| Tobias Haas | tobias.haas@desy.de | ZEUS, Spokesperson |
| Krzysztof Wrona | krzysztof.wrona@desy.de | ZEUS, Offline Coordinator |
| Janusz Szuba | janusz.szuba@desy.de | ZEUS, Offline Coordinator |
| Gunar Schnell | gunar.schnell@desy.de | HERMES / DESY Zeuthen |
| Takashi Sasaki | takashi.sasaki@kek.jp | KEK, Computing Centre |
| Nobu Katayama | nobu.katayama@kek.jp | Belle, Computing Coordinator |
| Fabio Hernandez | fabio.hernandez@cern.ch | CC-IN2P3, France |
| Salvatore Mele | Salvatore.Mele@cern.ch | CERN / PARSE |
| Andre Holzner | Andre.Georg.Holzner@cern.ch | CERN / LEP / L3 |
| Frederic Hemmer | frederic.hemmer@cern.ch | CERN / IT |
| Matthias Schroeder | matthias.schroder@cern.ch | CERN / IT |
| Olof Barring | olof.barring@cern.ch | CERN / IT |
| Rene Brun | rene.brun@cern.ch | CERN / ROOT |
| Marcello Maggi | marcello.maggi@cern.ch | CERN / ALEPH |
| Peter Igo-Kemenes | peter.igo-kemenes@cern.ch | CERN and Gjovik / PARSE |
| Jos Van Wezel | jos.vanwezel@kit.edu | GridKa |
| Andreas Heiss | andreas.heiss@kit.edu | GridKa |
| Gang Chen | Gang.Chen@ihep.ac.cn | IHEP Computing Centre |
| Yifang Wang | yfwang@ihep.ac.cn | IHEP BES III, Spokesperson |
| David Asner | asner@physics.carleton.ca | CLEO, Spokesperson |
| Daniel Riley | daniel.riley@cornell.edu | CLEO |
| David Corney | david.corney@stfc.ac.uk | Rutherford Lab / STFC |
| John Gordon | john.gordon@stfc.ac.uk | Rutherford Lab / STFC |

## Coordination

Chair: Cristinel Diaconu
Working Groups Convenors:
  Physics Case                     François Le Diberder
  Preservation Models        David South, Homer Neal
  Facilities and Technologies  Stephen Wolbers, Yves Kemp
  Governance                    Salvatore Mele

## International Steering Committee

DESY-IT: Volker Gülzow (DESY)
H1: Cristinel Diaconu (CPPM/DESY)
ZEUS: Tobias Haas (DESY)
FNAL/DoE: Amber Boehnlein (DoE)
FNAL-IT: Victoria White (FNAL)
D0: Dmitri Denisov (FNAL), Darien Wood (FNAL)
CDF: Jacobo Konigsberg (FNAL), Robert Roser (FNAL)
IHEP-IT: Gang Chen (IHEP)
BES III: Yifang Wang (IHEP)
KEK-IT: Takashi Sasaki (KEK)
Belle: Masanori Yamauchi (KEK), Tom Browder (Hawaii)
SLAC-IT: Richard Mount (SLAC)
BaBar: Francois Le Diberder (LAL/SLAC)
CERN-IT: Frederic Hemmer (CERN)
CERN/PARSE: Salvatore Mele (CERN)
CLEO: David Asner (Carleton)
STFC: John Gordon (RAL)

## International Advisory Committee

Chairs: Jonathan Dorfan (SLAC) and Siegfried Bethke (MPI Munich)
Gigi Rolandi (CERN), Michael Peskin (SLAC), Dominique Boutigny (IN2P3), Young-Kee Kim (FNAL), Hiroaki Aihara (IPMU/Tokyo)